\documentclass[11pt,a4paper,psamsfonts,twoside,reqno]{amsart}
\usepackage[latin1]{inputenc}
\usepackage{amssymb,amsthm}
\usepackage[mathscr]{eucal}
\usepackage{mathrsfs}
\usepackage{graphicx,pstricks}
\usepackage{scalefnt}

\theoremstyle{plain}

\theoremstyle{definition}

\theoremstyle{remark}

\begin{document}
\title{Generalized Heisenberg algebras and $k$-generalized Fibonacci numbers}
\author{Matthias Schork}
\address{Alexanderstr. 76\\ 60489 Frankfurt, Germany}
\email{mschork@member.ams.org}
\date{\today}
\abstract 
It is shown how some of the recent results of de Souza et al. \cite{souza} can be generalized to describe Hamiltonians whose eigenvalues are given as $k$-generalized Fibonacci numbers. Here $k$ is an arbitrary integer and the cases considered by de Souza et al. corespond to $k=2$.  
\endabstract
\maketitle

\section{Introduction}
Curado and Rego-Monteiro introduced in \cite{cur} a new algebraic structure generalizing the Heisenberg algebra and containing also the $q$-deformed oscillator as a particular case. This algebra, called {\it generalized Heisenberg algebra}, depends on an analytical function $f$ and the eigenvalues $\alpha_n$ of the Hamiltonian are given by the one-step recurrence $\alpha_{n+1}=f(\alpha_n)$. This structure has been used in different physical situations, see the references given in the recent paper \cite{souza}. In the same paper \cite{souza} de Souza et al. introduced an {\it extended two-step Heisenberg algebra} having many interesting properties. In particular, they showed that in certain special cases the eigenvalues of the involved Hamiltonian are given by the well-known Fibonacci numbers, i.e., satisfy a two-step recurrence. It is the aim of the present note to show how one may introduce for arbitrary natural numbers $k$ an {\it extended $k$-step Heisenberg algebra} which reduces for $k=2$ to the one discussed in \cite{souza} (and for $k=1$ to the one in \cite{cur}). In particular, the eigenvalues of the involved Hamiltonian are given in special cases by the $k$-generalized Fibonacci numbers \cite{miles}. For the convenience of the reader we now recall the structure of the extended two-step Heisenberg algebra, using the notations of \cite{souza}. It is generated by the set of operators $\{H,a^{\dag},a,J_3\}$ where $H=H^{\dag}$ is the Hamiltonian, $a$ and $a^{\dag}$ - with $a=(a^{\dag})^{\dag}$ - are the usual step operators and $J_3=J_3^{\dag}$ is an additional operator. These operators satisfy the following relations:
\begin{eqnarray}
Ha^{\dag}&=&a^{\dag}(f(H)+J_3), \hspace{0,5cm}\mbox{and} \hspace{0,5cm}aH=(f(H)+J_3)a,\label{com}\\
J_3a^{\dag}&=&a^{\dag}g(H),\hspace{0,5cm}\mbox{and}\hspace{0,5cm} aJ_3=g(H)a,\label{comm}\\ \left[a,a^{\dag}\right]&=&f(H)-H+J_3,\\ \left[H,J_3\right]&=&0.
\end{eqnarray}     
The functions $f$ and $g$ are assumed to be analytical (in fact, the case discussed most thoroughly is the one where these functions are linear). In the Fock space representation one has the normalized vacuum state $|0\rangle$ defined by the relations 
\[
a|0\rangle=0,\,\,H|0\rangle=\alpha_0|0\rangle,\,\, J_3|0\rangle=\beta_0|0\rangle,
\]
where $\alpha_0$ and $\beta_0$ are real numbers. By some algebra one can show the following consequences
\[a^{\dag}|n\rangle=N_n|n+1\rangle,\,\,
a|n\rangle=N_{n-1}|n-1\rangle,
H|n\rangle =\alpha_n|n\rangle,\,\,
J_3|n\rangle = \beta_n|n\rangle,
\]
(the first equation is the definition of the constants $N_n$) where the following relations hold true
\begin{eqnarray}
\alpha_{n+1}&=&f(\alpha_n)+\beta_n, \label{con1}\\
\beta_{n+1}&=&g(\alpha_n),\label{con2}\\
N_{n+1}^2&=&N_n^2+f(\alpha_{n+1})-\alpha_{n+1}+\beta_{n+1}.\label{con3}
\end{eqnarray}
Combining (\ref{con1}) and (\ref{con2}) shows the two-step dependence of the eigenvalues of the Hamiltonian (i.e., the energy-levels): $\alpha_{n+1}=f(\alpha_n)+g(\alpha_{n-1})$. In particular, choosing linear functions $f(x)=rx$ and $g(x)=sx$ (with $r,s\in \mathbb{R}\setminus \{0\}$) yields the recursion relation $\alpha_{n+1}=r\alpha_n+s\alpha_{n-1}$, showing that the energy-levels comprise a generalized Fibonacci sequence (which reduces in the special case $r=1=s$ to the usual Fibonacci sequence), thus providing a connection to Fibonacci-oscillators \cite{arik}. We may write the recursion relation of the linear case as follows:
\begin{equation}\label{matrix}
\left({\alpha_{n+1} \atop \beta_{n+1} }\right)=\left(\begin{array}{cc} r&1 \\ s & 0 \end{array} \right)\left({\alpha_{n} \atop \beta_{n} }\right).
\end{equation}
Using $\beta_{n+1}=s\alpha_n$, we can write this also in terms of the $\{\alpha_n\}_{n\in \mathbf{N}}$ only (and where we switch the order of the vector components):
\begin{equation}\label{matrix2}
\left({\alpha_{n} \atop \alpha_{n+1} }\right)=\left(\begin{array}{cc} 0&1\\ s & r \end{array} \right)\left({\alpha_{n-1} \atop \alpha_{n} }\right)\equiv T_2(r,s)\left({\alpha_{n-1} \atop \alpha_{n} }\right)= T_2(r,s)^n\left({\alpha_{-1} \atop \alpha_{0} }\right)
\end{equation}
where we have used the convention $\alpha_{-1}\equiv \frac{\beta_0}{s}$. In \cite{souza} the equation (\ref{matrix}) is studied from a stability analysis and the fixed points are studied. For this the eigenvalues $\lambda_{\pm}(r,s)=\frac{r\pm \sqrt{r^2+4s}}{2}$ of $T_2(r,s)$ are determined. However, it is also possible to use these to give an explicit expression for the $\alpha_n$ due to (\ref{matrix2}). The result can also be derived by {\it Binet's formula}. Note that the case $r=1=s$ of the classical Fibonacci numbers (where $\alpha_0=1$ and $\alpha_{-1}\equiv \beta_0=0$) gives rise to $\lambda_{\pm}(1,1)=\frac{1\pm \sqrt{5}}{2}$ where $\lambda_{+}(1,1)=\frac{1+\sqrt{5}}{2}$ is the {\it golden ratio} describing the limit $\lim_{n\rightarrow \infty} \frac{\alpha_{n+1}}{\alpha_n}$ of two consecutive elements of the sequence. One also has a relation between (\ref{matrix}) and (generalized) Fibonacci chains via {\it substitution rules} \cite{souza}. In the case $r=1=s$ one associates to (\ref{matrix}) the substitution rule $A \rightarrow AB$ and $B\rightarrow A$, thus yielding the Fibonacci chain $A\rightarrow AB\rightarrow ABA\rightarrow ABAAB\rightarrow \cdots$ whose length grows according to the Fibonacci numbers $1,2,3,5,\ldots$. In the case of arbitrary natural numbers $r$ and $s$ one has $r+1$ different possible substitution rules, namely $A\rightarrow A\cdots ABA\cdots A$ ($A$ appears $r$ times and $B$ appears at the $i$-th place with $1\leq i \leq r+1$) and $B\rightarrow A\cdots A$ ($s$ times). 

In this note we show how some results of de Souza et al. \cite{souza} - corresponding to the case $k=2$ - can be generalized to arbitrary natural numbers $k\geq 3$ where the integer $k$ denotes the order of the recursion relation which the eigenvalues of the Hamiltonian satisfy. As recalled above, the case $k=2$ is associated to the well-known Fibonacci numbers which have appeared in numerous physical applications, many of them in the context of quasicrystals (through substitution rules - see, e.g., \cite{hida,ilan} and the references therein as first steps into the vast literature). The case of larger $k$ is associated to the $k$-generalized Fibonacci numbers introduced by Miles \cite{miles} and discussed from many different points of view, see, e.g., \cite{hoggatt,fuller,phil,phil2,lev,dubeau,lee0,mourach,dubeau2,lee3,mourach2,lee}. Up to now these numbers have appeared only in few physical applications (in particular in connection with generalized exclusion statistics \cite{rach,ich} or the statistics of certain lattice animals which is closely connected to percolation problems \cite{turban,turban1}), although they give also rise to interesting substition rules as will be discussed later on. For the convenience of the reader we have treated the case $k=3$ seperately from the general case, although this has led to some repetitions. 
   
\section{The case $k=3$}
In this section we want to extend the structure discussed in the last section to an extended three-step Heisenberg algebra, i.e., to the case $k=3$. We will accomplish this by introducing a further operator $J_4$ with $J_4^{\dag}=J_4$ and a further analytical function $h$. The main idea is that the $n$-th eigenvalue of this new operator is connected to the $(n-2)$-th eigenvalue of $H$ (generalizing (\ref{con2})) and that the $(n+1)$-th eigenvalue of $H$ - i.e., $\alpha_{n+1}$ - is given by $f(\alpha_n) $ and the sum of the $n$-th eigenvalues of $J_3$ and $J_4$ (generalizing (\ref{con1})). Let us recall that (\ref{con2}) follows from (\ref{comm}) due to
\[
\beta_{n}|n\rangle=J_3|n\rangle=\frac{1}{N_{n-1}}J_3a^{\dag}|n-1\rangle \stackrel{(\ref{comm})}{=}\frac{1}{N_{n-1}}a^{\dag}g(H)|n-1\rangle=g(\alpha_{n-1})|n\rangle.
\]   
Thus, to obtain a connection between the $n$-th eigenvalue of $J_4$ and the $(n-2)$-th eigenvalue of $H$, we will have to consider what happens when $(a^{\dag})^2$ is interchanged with $J_4$. This motivates (\ref{komm33}). Similarly, to obtain the generalization of (\ref{con1}) we first recall that (\ref{con1}) depends on (\ref{com}) as follows
\[
\alpha_{n+1}|n+1\rangle=\frac{1}{N_{n}}Ha^{\dag}|n\rangle \stackrel{(\ref{com})}{=}\frac{1}{N_{n}}a^{\dag}(f(H)+J_3)|n\rangle=(f(\alpha_n)+\beta_n)|n\rangle.
\]
Thus, to realize the second part of the idea we have to generalize (\ref{con1}) by replacing on the right-hand side $J_3$ by $J_3+J_4$. This motivates (\ref{komm31}). Of course, it is then necessary to close the algebra in a consistent fashion. After these motivations we now introduce the {\it extended three-step Heisenberg algebra} through the set of operators $\{H,a^{\dag},a,J_3,J_4\}$ satisfying the commutation relations (\ref{komm31})-(\ref{komm36}):   
\begin{eqnarray}
Ha^{\dag}&=&a^{\dag}(f(H)+J_3+J_4),\label{komm31}\\
J_3a^{\dag}&=&a^{\dag}g(H),\label{komm32}\\
J_4(a^{\dag})^2&=&(a^{\dag})^2h(H), \label{komm33}\\ 
\left[a,a^{\dag}\right]&=&f(H)-H+J_3+J_4,\label{komm34}\\ 
\left[H,J_3\right]&=&0, \left[H,J_4\right]=0,\\
\left[J_3,J_4\right]&=&0.\label{komm36}
\end{eqnarray}     
(Here we have omitted the adjoint equations in (\ref{komm31})-(\ref{komm33}).) In the Fock space representation one has the normalized vacuum state $|0\rangle$ defined by the relations 
\[
a|0\rangle=0,\,\,
H|0\rangle=\alpha_0|0\rangle,\,\,
J_3|0\rangle=\beta_0|0\rangle,\,\, J_4|0\rangle=\gamma_0|0\rangle,
\]
where $\alpha_0, \beta_0$ and $\gamma_0$ are real numbers. The first normalization constant is shown to be $N_0^2=f(\alpha_0)-\alpha_0+\beta_0+\gamma_0$ (using (\ref{komm34})). By some algebra one can show the following consequences:
\begin{eqnarray*}
a^{\dag}|n\rangle&=&N_n|n+1\rangle,\,\,
a|n\rangle=N_{n-1}|n-1\rangle,\\
H|n\rangle &=&\alpha_n|n\rangle,\,\,
J_3|n\rangle = \beta_n|n\rangle, \,\,J_4|n\rangle = \gamma_n|n\rangle.
\end{eqnarray*}
It is now possible to show the following relations generalizing (\ref{con1}) and (\ref{con2})
\begin{eqnarray}
\alpha_{n+1}&=&f(\alpha_n)+\beta_n+\gamma_n, \label{con31}\\
\beta_{n+1}&=&g(\alpha_n),\label{con32}\\
\gamma_{n+1}&=&h(\alpha_{n-1}),\label{con33}\\
N_{n+1}^2&=&N_n^2+f(\alpha_{n+1})-\alpha_{n+1}+\beta_{n+1}+\gamma_{n+1}.\label{con34}
\end{eqnarray}
The new relation (\ref{con33}) is crucial for the following. The proof of it is analogous to (\ref{con32}):
\begin{eqnarray*}
\gamma_n|n\rangle &=&J_4|n\rangle = \frac{1}{N_{n-1}N_{n-2}}J_4(a^{\dag})^2|n-2\rangle \stackrel{(\ref{komm33})}{=}  \frac{1}{N_{n-1}N_{n-2}}(a^{\dag})^2h(H)|n-2\rangle \\ &=&\frac{1}{N_{n-1}}\frac{1}{N_{n-2}}(a^{\dag})^2h(\alpha_{n-2})|n-2\rangle =h(\alpha_{n-2})|n\rangle.
\end{eqnarray*}  
Thus, $\gamma_n=h(\alpha_{n-2})$. Taking (\ref{con31})-(\ref{con33}) together, we obtain the following relation 
\begin{equation}
\alpha_{n+1}=f(\alpha_n)+g(\alpha_{n-1})+h(\alpha_{n-2}),
\end{equation}
showing that the eigenvalues $\alpha_n$ of the Hamiltonian satisfy a three-step recurrence. Choosing linear functions $f(x)=rx, g(x)=sx$ and $h(x)=tx$ (with $r,s,t\in \mathbb{R}\setminus\{0\}$) yields the recursion relation 
\begin{equation}\label{rekur3}
\alpha_{n+1}=r\alpha_n+s\alpha_{n-1}+t\alpha_{n-2},
\end{equation}
showing that the energy-levels comprise a {\it generalized Tribonacci sequence} (and in the special case $r=s=t=1$ this reduces to the usual {\it Tribonacci sequence}). In analogy to the case $k=2$ considered above, we can write the relations (\ref{con31})-(\ref{con33}) in the linear case as follows:
\begin{equation}\label{matrixr3}
\left(\begin{array}{c}\alpha_{n+1} \\ \beta_{n+1} \\ \gamma_{n+1} \end{array}\right)
=\left(\begin{array}{ccc} r&1&1 \\ s & 0&0 \\0&\frac{t}{s}&0 \end{array} \right)
\left(\begin{array}{c}\alpha_{n} \\ \beta_{n} \\ \gamma_{n} \end{array}\right).
\end{equation}
Note that we have used here $\gamma_{n+1}=t\alpha_{n-1}$ together with $\beta_{n}=s\alpha_{n-1}$ to express $\gamma_{n+1}$ through $\beta_n$. As in the case $k=2$ we can bring this into a form where only the $\{\alpha_n\}_{n\in \mathbf{N}}$ are involved (again, the order of the vector components is switched):
\begin{equation}\label{matrixr32}
\left(\begin{array}{c}\alpha_{n-1} \\ \alpha_{n} \\ \alpha_{n+1} \end{array}\right)
=\left(\begin{array}{ccc} 0&1&0 \\ 0 & 0&1 \\t&s&r \end{array} \right)
\left(\begin{array}{c}\alpha_{n-2} \\ \alpha_{n-1} \\ \alpha_{n} \end{array}\right)
\equiv T_3(r,s,t)\left(\begin{array}{c}\alpha_{n-2} \\ \alpha_{n-1} \\ \alpha_{n} \end{array}\right).
\end{equation}
Introducing for notational convenience the vectors $\vec{\alpha}^{(n+1)}_3 := (\alpha_{n-1} \,\, \alpha_n \,\,\alpha_{n+1})^t$ and using the conventions $\alpha_{-1}\equiv\frac{\beta_{0}}{s}$ and $\alpha_{-2}\equiv\frac{\gamma_{0}}{t}$, we can iterate (\ref{matrixr32}) and obtain
\begin{equation}\label{matrixr33}
\vec{\alpha}^{(n+1)}_3=T_3(r,s,t)^n\vec{\alpha}^{(0)}_3.
\end{equation}
It is the clear that the eigenvalues $\lambda_1(r,s,t),\lambda_2(r,s,t),\lambda_3(r,s,t)$ of $T_3(r,s,t)$ will play an important role for the ``dynamics'' of the system (i.e., the solutions of (\ref{rekur3})) - as in the case $k=2$ considered above; here we have assumed that the parameters $r,s,t$ are chosen appropriately such that three eigenvalues exist. One also has a {\it generalized Binet's formula} (see, e.g., \cite{miles,lev,lee3,lee}) which allows one to express the general solution $\alpha_n$ through the eigenvalues $\lambda_i(r,s,t)$ and the initial values $\alpha_0,\beta_0,\gamma_0$. However, note that a physical requirement will be that all energy-levels are greater than zero, i.e., $E_n\equiv \alpha_n \geq 0$. This restricts the set of possible functions $f,g,h$, and, in the particular case of linear functions it restricts the coefficients $r,s,t$. For example, if $r,s,t \geq 0$ than the energy-levels build a nondecreasing sequence, i.e., $E_{n+1}\geq E_n $. This seems to be a very natural condition (otherwise the counting seems to be very strange).

We can generalize the connection between (\ref{matrix}) and (generalized) Fibonacci chains via substitution rules described at the end of the first section as follows. Starting from (\ref{matrixr3}), we introduce the alphabet $\{A,B,C\}$ and assume - in analogy to the case $k=2$ - that $r,s,t$ are natural numbers; in addition, we furthermore assume that $\frac{t}{s}$ is a natural number, i.e., $t$ is a multiple of $s$. Then there exist several substitution rules associated to (\ref{matrixr3}):
\begin{equation}\label{substr3}
A\rightarrow A^{l_1} A_1 A^{l_2}A_2A^{l_{3}}, \hspace{0,5cm} B\rightarrow A^s,\hspace{0,5cm} C\rightarrow B^{\frac{t}{s}},
\end{equation}
where $(A_1,A_2)$ is a permutation of $(B,C)$, $0\leq l_i\leq r$ for $1\leq i \leq 3$ and $l_1+l_2+l_3=r$ (here we have denoted the concatenation of $p$ letters $A$ by $A^p=A\cdots A$ ($p$ times)). The number of substitution rules is given by $(r+1)(r+2)$ which is the number of different words of length $r+2$ in the letters $\{A,B,C\}$ where the letter $A$ appears exactly $r$ times. As an example, let $r=2,s=1,t=2$. A possible substitution rule is then given by $A\rightarrow ABAC, B\rightarrow A, C\rightarrow BB$. This generates the following generalized Fibonacci chain: $A\rightarrow ABAC \rightarrow ABACAABACBB\rightarrow ABACAABACBBABACABACAABACBBAA\rightarrow...$. It is evident that the lengths grow very rapidly ($1,4,11,29,\ldots$).

\section{The generalization to arbitrary $k\geq 2$}  
In this section we will generalize the situation of the preceeding sections to the case of arbitrary $k\geq 2$. To obtain a transparent formulation we will change the notation to a more convenient one. In addition to the Hamiltonian $H$ and the step operators $a$ and $a^{\dag}$ we have the operators $J_i$ with $J_i^{\dag}=J_i$ for $i=2,\ldots,k$ and associated analytic functions $f_i$ for $1\leq i \leq k$. It is clear from the explicit discussion of the case $k=3$ that the main step consists in introducing new commuation relations for $J_i$ and $(a^{\dag})^{i-1}$. Thus, the set of operators of the {\it extended $k$-step Heisenberg algebra} consists of $\{H,a^{\dag},a,J_2,J_3,\ldots,J_k\}$ and the commutation relations satisfied by these operators are given by (\ref{kommr1})-(\ref{kommr5}):
\begin{eqnarray}
Ha^{\dag}&=&a^{\dag}\left(f_1(H)+\sum_{i=2}^kJ_i\right),\label{kommr1}\\
J_i(a^{\dag})^{i-1}&=&(a^{\dag})^{i-1}f_{i}(H) \,\,\mbox{for}\,\, 2 \leq i \leq k,\label{kommr2}\\ 
\left[a,a^{\dag}\right]&=&f_1(H)-H+\sum_{i=2}^kJ_i,\\ 
\left[H,J_i\right]&=&0 \,\,\mbox{for}\,\, 2 \leq i \leq k,\\
\left[J_i,J_j\right]&=&0 \,\,\mbox{for}\,\, 2 \leq i < j \leq k.\label{kommr5}
\end{eqnarray} 
(Here the adjoint equations in (\ref{kommr1}) and (\ref{kommr2}) have been omitted.) Clearly, choosing $k=3$ reproduces the situation considered in the last section (if one also makes the replacement $(J_2,J_3,f_1,f_2,f_3)\rightsquigarrow(J_3,J_4,f,g,h)$). In the Fock space representation one has the normalized vacuum state $|0\rangle$ defined by the relations 
\[
a|0\rangle=0,\,\, H|0\rangle=\alpha_0^{(1)}|0\rangle,\,\,
J_i|0\rangle=\alpha^{(i)}_0|0\rangle \,\,\mbox{for}\,\, 2 \leq i \leq k,
\]
where $\alpha_0^{(i)}$ for $1\leq i\leq k$ are real numbers. It follows that
\begin{eqnarray*}
a^{\dag}|n\rangle&=&N_n|n+1\rangle,\,\,
a|n\rangle=N_{n-1}|n-1\rangle,\\
H|n\rangle &=&\alpha_n^{(1)}|n\rangle,\,\,
J_i|n\rangle = \alpha^{(i)}_n|n\rangle\,\,\mbox{for}\,\, 2 \leq i \leq k.
\end{eqnarray*}
The algebraic relations imply
\begin{eqnarray}
\alpha_{n+1}^{(1)}&=&f_1(\alpha_n^{(1)})+\sum_{i=2}^k\alpha_n^{(i)}, \label{conr1}\\
\alpha^{(i)}_{n+1}&=&f_i(\alpha_{n-i+2}^{(1)})\,\,\mbox{for}\,\, 2 \leq i \leq k, \label{conr2}\\
N_{n+1}^2&=&N_n^2+f_1(\alpha_{n+1}^{(1)})-\alpha_{n+1}^{(1)}+\sum_{i=2}^k \alpha^{(i)}_{n+1}.\label{conr3}
\end{eqnarray}
As in the case $k=3$ the relations (\ref{conr2}) are crucial for the following and they are shown in the same way as (\ref{con33}) by application of (\ref{kommr2}):
\[
\alpha^{(i)}_n|n\rangle =\frac{J_i(a^{\dag})^{i-1}}{\kappa(n,i)}|n-i+1\rangle \stackrel{(\ref{kommr2})}{=}  \frac{(a^{\dag})^{i-1}f_i(H)}{\kappa(n,i)}|n-i+1\rangle=f_i(\alpha_{n-i+1}^{(1)})|n\rangle.
\]  
(Here we have denoted by $\kappa(n,i)=N_{n-1}N_{n-2}\cdots N_{n-i+1}$ the product of the constants appearing in the intermediate steps of the calculation.) Taking together (\ref{conr1}) and (\ref{conr2}) yields the relation 
\begin{equation}
\alpha_{n+1}^{(1)}=f_1(\alpha_n^{(1)})+\sum_{i=2}^kf_i(\alpha_{n-i+1}^{(1)}).\label{recurr}
\end{equation}
Let us now assume that the functions are linear, i.e., $f_i(x)=\lambda_ix$ for $1\leq i \leq k$ (with $\lambda_i \in \mathbb{R}\setminus\{0\}$). Then (\ref{recurr}) reduces to
\begin{equation}\label{rekurr}
\alpha_{n+1}^{(1)}=\lambda_1\alpha_n^{(1)}+\lambda_2\alpha^{(1)}_{n-1}+\cdots + \lambda_k\alpha^{(1)}_{n-k+1},
\end{equation}
i.e., to the recursion relation of the $k$-{\it generalized Fibonacci numbers} \cite{miles,lev}. In analogy to the cases $k=2,3$ considered above we can write the relations (\ref{conr1}) and (\ref{conr2}) in the linear case as matrix equation. For this we first observe that one has for $2\leq i \leq k$ that $\alpha^{(i)}_{n+1}=\lambda_i \alpha_{n-i+2}^{(1)}$. Since on the right-hand side only expressions with index $n$ should appear we have to reexpress $\alpha_{n-i+2}^{(1)}$. However, one finds that $\alpha^{(i-1)}_{n}=\lambda_{i-1} \alpha_{n-i+2}^{(1)}$, implying the sought-for relation $\alpha^{(i)}_{n+1}=\frac{\lambda_i}{\lambda_{i-1}}\alpha^{(i-1)}_{n}$. To bring the recursion (\ref{rekurr}) into the desired form we have to notice that $\alpha^{(m)}_n=\lambda_m\alpha^{(1)}_{n-m+1}$ so that this equation can be written as $\alpha_{n+1}^{(1)}=\lambda_1\alpha_{n}^{(1)}+\alpha_n^{(2)}+\cdots+\alpha_n^{(k)}.$ The resulting equation is
\begin{equation}\label{matrixrr}
\left(\begin{array}{c}\alpha_{n+1}^{(1)} \\ \alpha_{n+1}^{(2)} \\\alpha_{n+1}^{(3)} \\ \alpha_{n+1}^{(4)} \\ \vdots  \\ \alpha_{n+1}^{(k)}  \end{array}\right)
=\left(\begin{array}{cccccc} \lambda_1&1 & 1&1&\cdots&1\\ \lambda_2&0 & 0&0&\cdots&0 \\ 0& \frac{\lambda_3}{\lambda_2} & 0&0&\cdots&0 \\ 0&0 & \frac{\lambda_4}{\lambda_3}&0&\cdots&0\\\vdots &\ddots &\ddots &\ddots &\ddots&\vdots \\0&0&0&0&\frac{\lambda_k}{\lambda_{k-1}}&0 \end{array} \right)
\left(\begin{array}{c}\alpha_{n}^{(1)} \\ \alpha_{n}^{(2)} \\ \alpha_{n}^{(3)} \\ \alpha_{n}^{(4)} \\ \vdots  \\ \alpha_{n}^{(k)}  \end{array}\right).
\end{equation}
As in the cases $k=2$ and $k=3$ we can bring this into a form where only the $\{\alpha_n^{(1)}\}_{n\in \mathbf{N}}$ are involved:
\begin{equation}\label{matrixrr2}
\left(\begin{array}{c}\alpha_{n-k+2}^{(1)} \\ \alpha_{n-k+3}^{(1)} \\ \vdots  \\ \alpha_{n-1}^{(1)} \\  \alpha_{n}^{(1)} \\ \alpha_{n+1}^{(1)}  \end{array}\right)
=\left(\begin{array}{cccccc} 0&1 &0&\cdots  &\cdots &0 \\ 
0&0& 1&\ddots&&\vdots \\ 
\vdots & \vdots &\ddots& \ddots &\ddots& \vdots \\ 
0&0&\cdots &0 &1&0\\ 
0 &0&\cdots &0 &0&1\\ 
\lambda_k&\lambda_{k-1} &\cdots &\lambda_3&\lambda_2&\lambda_1 \end{array} \right)
\left(\begin{array}{c}\alpha_{n-k+1}^{(1)} \\ \alpha_{n-k+2}^{(1)} \\ \vdots  \\ \alpha_{n-2}^{(1)} \\  \alpha_{n-1}^{(1)} \\ \alpha_{n}^{(1)}  \end{array}\right).
\end{equation}
Let us denote the matrix involved by $T_k(\lambda)$; here we abbreviate the coefficients $\lambda_i$ as a vector $\lambda=(\lambda_1,\ldots,\lambda_k)$. Introducing for notational convenience the vectors $\vec{\alpha}^{(n+1)}_k := (\alpha_{n-k+2}^{(1)} \,\, \cdots \,\,\alpha_{n}^{(1)}\,\, \alpha_{n+1}^{(1)})^t$ and using the convention $\alpha_{-m}^{(1)}=\frac{\alpha_{0}^{(m+1)}}{\lambda_{m+1}}$ we can iterate (\ref{matrixrr2}) and obtain the generalization of (\ref{matrixr33}):
\begin{equation}\label{matrixrr3}
\vec{\alpha}^{(n+1)}_k=T_k(\lambda)^n\vec{\alpha}^{(0)}_k.
\end{equation}
It is the clear that the eigenvalues of $T_k(\lambda)$ will play an important role for the ``dynamics'' of the system; however, the physical requirement of positive energies $E_n\equiv \alpha_n^{(1)}\geq 0$ will restrict the set of possible coefficients $\lambda_i$ as in the case $k=3$ discussed above. One also has a {\it generalized Binet's formula} (see, e.g., \cite{miles,lev,lee3,lee}) which allows one to express the general solution $\alpha_n^{(1)}$ through the eigenvalues of $T_k(\lambda)$ and the initial values $\alpha_0^{(i)}$. Let us mention that the matrix $T_k(\lambda)$ is called in the mathematical literature (in particular in the case where all $\lambda_i=1$) the $k$-{\it generalized Fibonacci-matrix} $Q_k$ \cite{lee,lee3,mourach,mourach2}. If the coefficients $\lambda_i$ satisfy $\lambda_i\geq 0$ and $\lambda_1+\cdots+\lambda_k=1$ then $T_k(\lambda)$ is a stochastic matrix and the recursion relation (\ref{matrixrr3}) describes a Markov chain \cite{mourach,mourach2}. If we denote the $k$-generalized Fibonacci numbers introduced by Miles \cite{miles} by $F_n^{(k)}$, then one has in the case where all $\lambda_i=1$ and where the initial values are given by $\alpha_0^{(1)}=1$ as well as $\alpha_0^{(i)}=0$ for $2\leq i \leq k$ the following relation
\begin{equation}
E_n\equiv \alpha_n^{(1)}=F_{n+k-1}^{(k)}=\sum_{{ 0\leq a_1,\ldots,a_k\leq n+k-1}\atop {a_1+2 a_2+\cdots +
ka_k=n}}\frac{(a_1+\cdots+a_k)!}{a_1!\cdots a_k!}.
\end{equation}
Here we have used the explicit expression for the $k$-generalized Fibonacci numbers given by \cite{miles}:
\[
F_m^{(k)}=\sum_{{ 0\leq a_1,\ldots, a_k\leq m}\atop {a_1+2 a_2+\cdots +
ka_k=m-k+1}}\frac{(a_1+\cdots+a_k)!}{a_1!\cdots a_k!}.
\]
The case with arbitrary $\lambda_i$ as well as arbitrary initial values can be treated in a similar form using \cite{lev}.

Clearly, it is again possible to associate substitution rules to (\ref{matrixrr}) by introducing an alphabet $\{A_1,A_2,\ldots,A_k\}$. For this we have to assume that all $\lambda_i$ with $1\leq i\leq k$ as well as all quotients $q_i:=\frac{\lambda_i}{\lambda_{i-1}}$ with $1\leq i\leq k-1$ are natural numbers. Note that this implies $\lambda_3=q_3\lambda_2,\lambda_4=q_4\lambda_3=q_4q_3\lambda_2$ and in general $\lambda_m=q_mq_{m-1}\cdots q_3\lambda_2$. The appropriate substitution rules generalizing (\ref{substr3}) are
\begin{equation}
A_1\rightarrow A_1^{l_1} {A_{i_1}}A_1^{l_2} {A_{i_2}}\cdots {A_{i_{k-1}}}A_1^{l_{k}}, \hspace{0,1cm} A_2\rightarrow A_1^{\lambda_2}, \hspace{0,1cm} A_i\rightarrow A_{i-1}^{q_i}\hspace{0,2cm} \mbox{for } 3\leq i \leq k
\end{equation}
where $(i_1,i_2,\ldots,i_{k-1})$ is a permutation of $(2,3,\ldots,k)$, $0\leq l_i\leq \lambda_1$ for $1\leq i \leq k$ and $l_1+l_2+\cdots+l_k=\lambda_1$. The number of substitution rules is given by $(\lambda_1+1)(\lambda_1+2)\cdots (\lambda_1+k-1)$ which is the number of different words of length $\lambda_1+k-1$ in the letters $\{A_1,\ldots,A_k\}$ where the letter $A_1$ appears exactly $\lambda_1$ times.

\section{Conclusions}
In this note we have shown how some results of de Souza et al. \cite{souza} can be generalized from the case $k=2$ to arbitrary natural numbers $k\geq 3$ where the integer $k$ denotes the order of the recursion relation which the eigenvalues of the Hamiltonian satisfy. The case $k=2$ is associated to the well-known Fibonacci numbers which have appeared in numerous physical applications, many of them in the context of quasicrystals via substitution rules. The case of larger $k$ is associated to the $k$-generalized Fibonacci numbers introduced by Miles. It was pointed out that the case $k\geq 3$ gives also rise to interesting substitution rules which might be interesting for the study of quasicrystals.

\end{document}